\documentclass{article}

\usepackage{arxiv}

\usepackage[utf8]{inputenc} % allow utf-8 input
\usepackage[T1]{fontenc}    % use 8-bit T1 fonts
\usepackage{hyperref}       % hyperlinks
\usepackage{url}            % simple URL typesetting
\usepackage{booktabs}       % professional-quality tables
\usepackage{amsfonts}       % blackboard math symbols
\usepackage{nicefrac}       % compact symbols for 1/2, etc.
\usepackage{microtype}      % microtypography
\usepackage{lipsum}
\usepackage{graphicx}
\graphicspath{ {./images/} }

    \usepackage{epsfig,xspace}
    \usepackage{subcaption}
    \usepackage{calc}
    \usepackage{amssymb}
    \usepackage{amstext}
    \usepackage{amsmath}
    \usepackage{amsthm}
    \usepackage{booktabs} % For formal tables
    \usepackage{multicol}

    \usepackage[utf8]{inputenc}
    \usepackage[english]{babel}
    \usepackage{caption}
    \usepackage{hyperref}
    \usepackage{nicefrac} 
    \usepackage{pdflscape}
    \usepackage{array,multirow,graphicx}
    \usepackage{pifont}
    \newcommand{\cmark}{\ding{51}}
    \newcommand{\xmark}{\ding{55}}
    \newtheorem*{definition}{Definition}
    \newcommand{\remove}[1]{}
    \usepackage{xcolor}

    \newcommand{\LTR}{\textsf{LTR}\xspace}%{LTR\xspace}
    \newcommand{\FWLTR}{\textsf{FLTR}\xspace}%{FwLTR\xspace}
    \newcommand{\FwLTR}{\textsf{FLTR}\xspace}%{FwLTR\xspace}
    \newcommand{\FLTR}{\textsf{FLTR}\xspace}
    \newcommand{\ICR}{\textsf{ICR}\xspace}%{ICR\xspace}
    \newcommand{\Pagerank}{\textsf{PgR}\xspace}%{PageRank\xspace}
    \newcommand{\PageRank}{\textsf{PgR}\xspace}%{PageRank\xspace}
    \newcommand{\PR}{\textsf{PgR}\xspace}%{PageRank\xspace}
    \newcommand{\Betweenness}{\textsf{Btwn}\xspace}
    
    \newcommand{\gini}{\textsf{Gini}\xspace}
    \newcommand{\Gini}{\textsf{Gini}\xspace}
    \newcommand{\topTEN}{\textsf{Top10}\xspace}
    \newcommand{\topTENa}{\textsf{Top10\%Actors}\xspace}
    \newcommand{\topTENv}{\textsf{Top10\%Values}\xspace}
    \newcommand{\Amazon}{\textsf{Amazon}\xspace}
    \newcommand{\ArXiv}{\textsf{ArXiv}\xspace}
    \newcommand{\Caida}{\textsf{Caida}\xspace}
    \newcommand{\ENRON}{\textsf{ENRON}\xspace}
    \newcommand{\Epinions}{\textsf{Epinions}\xspace}
    \newcommand{\Gnutella}{\textsf{Gnutella}\xspace}
    \newcommand{\Higgs}{\textsf{Higgs}\xspace}
    
    \newcommand{\Wikipedia}{\textsf{Wikipedia}\xspace}

\title{Playing with Thresholds on the Forward Linear Threshold Rank}
\author{
    Maria J.~Blesa \\
    Computer Science Department \\ 
    Universitat Politècnica de Catalunya \\
    08034 Barcelona \\
    \texttt{maria.j.blesa@upc.edu} \\
    \texttt{\textsc{orcid:~}{0000-0001-8246-9926}} \\
    \And
    María Serna \\
    Computer Science Department and IMTech \\ 
    Universitat Politècnica de Catalunya \\
    08034 Barcelona, Spain \\
    \texttt{maria.serna@upc.edu} \\
    \texttt{\textsc{orcid:~}{0000-0001-9729-8648}}
}

\begin{document}
\maketitle
\begin{abstract}
    Social networks are the natural space for the  spreading of information and influence and have become a media themselves. Several models capturing that diffusion process have been proposed, most of them based on the Independent Cascade (IC) model or on the Linear Threshold (LT) model. The IC model is probabilistic while the LT model relies on the knowledge of an actor to be convinced, reflected in an associated individual threshold.  
    Although the LT-based models contemplate an individual threshold for each actor in the network, the existing studies so far have always considered a threshold of 0.5 equal in all actors (i.e., a simple majority activation criterion).
    Our main objective in this work is to start the study on how the dissemination of information on networks behaves when we consider other options for setting those thresholds and how many network actors end up being influenced by this dissemination. For doing so, we consider a recently introduced centrality measure based on the LT model, the Forward Linear Threshold Rank (\FLTR). We analyze experimentally the ranking properties for several networks in which the influence resistance threshold follows different schemes. Here we consider three different schemes: (1) uniform, in which all players have the same value; (2) random, where each player is assigned a threshold u.a.r. in a prescribed interval; and (3) determined by the value of another centrality measure on the actor. Our results show that the selection has a clear impact on the ranking, even quite significant and abrupt in some cases. We conclude that the social networks ranks that provide the best assignments for the individual thresholds are PageRank and \FLTR.
\end{abstract}

\begin{ack}
     M.~Blesa and M.~Serna acknowledge support by  MICIN/AEI/10.13039/501100011033 under grant PID2020-112581GB-C21 (MOTION).
\end{ack}

\newpage

% keywords can be removed

\section{Introduction}
\label{sec:introduction}
Nowadays the importance of social networks as a marketing tool is growing rapidly and spanning through diverse social networks and objectives. Think for example about viral marketing, that seeks to spread information about a product or service from person to person by word of mouth or sharing via the internet or email. The goal of viral marketing is to inspire individuals to share a marketing message (or opinion) to friends, family, and other individuals to create exponential growth in the number of its recipients. The so-called \emph{spread of influence} process is used here with a very specific goal and needs to be maximized until it stabilizes and no additional convictions are possible. How to select the initial group of individuals that start the spreading process is a key issue. 

Traditional social networks are labelled graphs in which the nodes represent individuals and the edges represent  weighted relations between them. There have been several proposal for modeling the spread of influence in a network. Among them, the \emph{Independent Cascade} (IC) model is a stochastic model proposed in \cite{Goldenberg}. It is based on the assumption that  whenever a node is activated, it will do (stochastically) attempt to activate a neighbor. 
The Linear Threshold (LT) model is a 
deterministic model for influence spread based on some ideas of collective behavior \cite{Kempe03}. In the LT model the strength of the tie between every pair of actors quantifies the capacity of one to influence the other and, 
additionally, each actor opposes a resistance to be influenced. A node gets influenced when  their active predecessors can exert enough influence to overpass its resistance. The node resistance is quantified by a threshold that quantifies the minimum required percentage of the total income weight needed to convince the actor. 
Several computational problems have been proposed that attempt to solve the problem of selecting the initial set of participants to spread the product in the most effective. The two most relevant being  the \emph{influence maximization}~\cite{Kempe03} and the \emph{target set selection problems}~\cite{Chen2009}.  Another line of research uses the spread of influence process to define centrality measures~\cite{Centrality,BGS2021}.

Trying to experimentally evaluate centrality measures based on the linear threshold model, we came with a lack of suitable data sets. Most of the social networks you can find in repositories do not include any kind on information about the resistance of a participant to be influenced.  Therefore, as no information about suitable threshold values to run the linear threshold model was collected, a single majority rule was assumed.  However this rule might not be a realistic one in many scenarios. In this paper, we attempt to shed some light on the effect that the selection of threshold has on one of the centrality measures based on the LT model. We focus our analysis in the \FwLTR rank which measures the influence that a node together with its immediate successors can exert~\cite{BGS2021}. 

We performed a series of experiments in order to assess how the threshold values affect the \FwLTR. We considered a subset of the social networks used in~\cite{Centrality}. To evaluate the effect of the threshold selection, we fix the method  to assign the influence thresholds and compute the \FLTR ranking. Over the ranking, we compute the traditional statistics and then we select some relevant sets of participants, the ten  in the top of the ranking, the 10\% in the top, and the participants whose rank is in the 10\% of the highest values, and compute the number of nodes influenced by those sets.  

We run three kind of experiments. In the first one we assign to each node a common threshold $\theta$. This type of uniform assignment was the usual scheme used in the literature until now, with $\theta=0.5$ to implement a simple majority rule. However, we try different values for $\theta$ others than 0.5. In the second experiment we assign random thresholds in different ranges. In the third experiment we follow a completely different schema and use the value of another centrality measure on the node to determine its threshold. 

Besides more specific comments given later, we comment here on the main discovered trends. Our first set of experiments clearly show  that, in most of the treated networks, the
range of threshold values more relevant to the maximization of the influence is within $[0.2, 0.5]$, where it is possible to accumulate greater proportion of the total influence diffused in groups of actors of small size. 
Interestingly enough, we found that a random distribution of the thresholds behaves practically the same with respect to our measures for the diffusion of influence whether they are in $(0, 0.5]$ or $(0, 1]$. This suggests that the fact that a minority fraction of the actors have high thresholds affects little in the actor's ability to influence in general.
Finally, we can observe in our last experiment that \Pagerank and \FLTR centralities provide the best thresholds, in the sense that they generate favorable conditions for a great influence expansion even when starting with reduced sets. In nearly all cases, it is enough to have the first 10 actors of the ranking to reach the entire network or its vast majority. 

\section{Centrality measures}
\label{sec:measures}

%\label{centr}
%\label{sec:centralitymeasures}

We outline the centrality measures used in different parts of this paper. As we will see, some of them are based on the relevance and the topological properties of the nodes, e.g., PageRank, while others focus on their influence over the network, e.g., the Linear Threshold Rank and the Independent Cascade Rank. We start describing the topology-based centrality measures, which were the ones used traditionally as centrality measures, and continue describing the influence-based ones that have been proposed more   recently.  As usual, we assume that the social network is represented by a directed graph $G=(V,E)$.

%an edge-weighted directed graph $G = (V, E, w)$ representing the social network is given.

\remove{
    %\medskip
    %\noindent
    \paragraph{Degree $-$}
    This is one of the simplest centrality measures. It consists on assigning the centrality based of the degree of the node. Given a directed graph $G = (V, E)$, for every node $i \in V$ , the degree centrality of $i$ is the degree itself of $i$, normalized by the number of nodes minus one.
    \[
    \mathrm{Degree}(i) = \frac{\delta_{i}}{n-1}
    \]
    where $\delta_{i}$ is the degree of $i$.
    
    %\medskip
    %\noindent
    \paragraph{Closeness $-$}
    For the closeness measure, a node has a higher centrality if it is close to the other nodes, the closer it is, the higher its rank will be. Given a directed graph $G = (V, E)$, for every node $i \in V$ , the closeness is the reciprocal sum of the distances from $i$ to the rest of the nodes in the graph.
    \[
    \mathrm{Closeness}(i) = \frac{nr-1}{n-1} \frac{nr-1}{\sum_{j \in V}^{} d(i,j)}
    \]
    where $nr$ is the number of accessible nodes from $i$ and $d(i, j)$ is the length of the shortest path between $i$ and $j$.
}

%\medskip
%\noindent
\paragraph{Betweenness $-$}
In the betweenness centrality, a node is more important if it belongs to the shortest path between any pair of nodes in the graph~\cite{Freeman77}.  For every node $i \in V$ , define \remove{we define the betweenness as the sum, for  $s,t \in V$, of the proportion of the number of shortest paths between $s,t$ that go through $i$ with respect to the number of shortest path between $s$ and $t$, i.e.,} 
\[
\Betweenness(i) = \sum_{s,t\in V-\{i\}}^{} \frac{\sigma _{st}(i)}{\sigma _{st}}
\]
where $\sigma _{st}$ is the number of shortest paths between $s$ y $t$, and $\sigma _{st}(i)$ the number of such shortest paths that pass by $i$.

\remove{
    %\medskip
    %\noindent
    \paragraph{Katz $-$}
    The Katz centrality~\cite{Katz} is a generalization of degree 
    centrality and it can also be viewed as a variant of eigenvector centrality. 
    While degree centrality measures the number of direct neighbors, the Katz 
    centrality measures the number of all nodes that can be connected through a 
    path, while the contributions of distant nodes are penalized. It is based on 
    the idea that an actor is important if it is linked to other important actors 
    or if it is highly linked. It overcomes the limitations of the eigenvector centrality when the graph has nodes that reach strongly connected components, but those connected components do not reach the node, which may occur in social networks. 
    
    Let $A$ be the adjacency matrix of the directed graph $G = (V, E)$ (i.e., $a_{ij}=1$ if there is an edge between $i$ y $j$, and a $a_{ij}=0$ otherwise). The $Katz(i)$ is defined as
    \[
    \mathrm{Katz}(i) = \alpha \sum_{j \in V}^{} a_{ji} \; \mathrm{Katz}(j) + \beta
    \]
    where $\beta$ is a constant independent from the structure of the network and $\alpha \in [0,\ldots,{\lambda^{-1}_{max}}]$ is the damping factor, being $\lambda_{max}$ the highest eigenvalue in $A$.
}

%\medskip
%\noindent
\paragraph{PageRank $-$}
One of the most popular centrality measures is the PageRank~\cite{Page}, that Google uses to assign relevance to web pages. A web page is more relevant  if other important web pages point to it. It uses a parameter $\alpha\in (0,1]$, that represents the probability that a user keeps jumping from a web page to another through the links that are between them (and thus, $1-\alpha$ represents the probability that the user goes to a random web page). Let $A$ be the adjacency matrix of $G$ (i.e., $a_{ij}=1$ if $(i,j)\in E$, and 0 otherwise), the PageRank (PR) of $i$ is given by
\[
\PR(i) = (1-\alpha) + \alpha \sum_{j \in V}^{} \frac{a_{ji} \; \PR(j)}{\delta ^{+}(j)}
\]
where $\delta^{+}(i)$ is the out degree of $i \in V$.

\bigskip
Perhaps the two most prevalent 
diffusion models in computer science are the \emph{Independent Cascade} 
model~\cite{Goldenberg} and the \emph{Linear Threshold} model~\cite{Kempe03} 
(see also~\cite{Shakarian15}). We define their corresponding influence-based centrality measures:

%\medskip
%\noindent
\paragraph{Independent Cascade Rank $-$}
The Independent Cascade Rank~\cite{Kempe05} is an influence-based centrality measure based on the Independent Cascade (IC) Model~\cite{Goldenberg}, which is a stochastic model.\remove{ that was initially proposed in the context of marketing.} It is based on the assumption that whenever a node is activated, it will (stochastically) do attempt to activate 
each actor he targets. Given an activated node $i\in V$, any neighbor $j$ such that $(i,j)\in E$ will be activated with a probability $p_{ij}$. When a new actor is 
activated, the process is repeated for this actor. The whole process ends when 
there are no active nodes with a new chance to spread its influence.  

Given an initial core $X\subseteq V$ and a probability $p\in [0,1]$ (where 
$\forall (i,j)\in E:p_{ij}=p$), the influence spread of $X$ is denoted by 
$F'(X,p)$. The Independent Cascade Rank 
of a node $u \in V$ is then defined as:
\[\ICR(u,p) = 
\frac{|F'(u,p)|}{\max_{v \in V} \{|F'(v,p)| \}}.\]

%\medskip
%\noindent
\paragraph{Linear Threshold Rank $-$}
The Linear Threshold Rank~\cite{Centrality} is also an influence-based centrality measure, based on the Linear Threshold (LT) Model~\cite{Kempe03}. Every node has an influence threshold, which represents the resistance of this node to be influenced by others. In the model, every edge $(u,v)$ also has a weight representing the influence that node $u$ has over node $v$. In practice, that weight is set to one since it is very different to quantify that concept.  

The influence algorithm starts with an initial predefined set of activated nodes. At every iteration, the active nodes will influence their neighbors. When the total influence that a node receives exceeds its influence threshold, then this node will become active and join the set of active nodes. 
%As long as new nodes join the set of active nodes, the spread of influence is still on progress.
The algorithm stops when the set of active nodes converges, i.e., when no new nodes are influenced. 
%In order to formally define the Linear Threshold Rank, we need to introduce the following concepts:

Given an initial set of active nodes $X\subseteq V$ and a threshold assignment  $\theta: V\to \mathbb{N}$, let 
\remove{
    \begin{definition} 
    An $\textit{influence graph}$ is a tuple $(G,w,f)$, where $G = (V,E)$ is a directed graph made by a set of actors $V$ and a set of relations $E; w : E \rightarrow N$ is a weight function that assigns a weight to each edge, representing the influence of one node to the other; $f : V \rightarrow N$ is a labeling function that quantifies how resistant to influence every node is. 
    \end{definition}
    
    %\begin{definition}
    %Given an influence graph $(G, w, f)$ and an initial active set $X \subseteq V$,
} 
$F_{t}(X) \subseteq V$ denote the set of activated nodes at the $t$-th iteration of the spreading process.  
%\end{definition}
At the first step ($t = 0$) only the nodes in $X$ are active, which means that $F_{0}(X) = X$. At the $t+1$ iteration, a node $i$ will be activated if, and only if, 
the sum of all the weights of the edges $\{i,j\}$, where $j$ is already active, is higher than the resistance (or influence threshold) of $i$, i.e., 
$
\sum_{j \subseteq F_{t}(X)}^{} w_{ij} \geq \theta(i),
$
In practice, it is usual to consider that all the edge weights are equal to one and thus a  node $i$ will be activated if, and only if, 
$$
\frac{\mid F_{t}(X) \cap \mathcal{N}(i)\mid}{|\mathcal{N}(i)|}  \geq \theta(i),
$$
where, $\mathcal{N}(u)=\{v \mid (u,v)\in E \lor (v,u)\in E\}$.
Observe, that the process is monotonic, therefore it stops after at most $n=|V|$ steps. Thus can define the spread of $X$ as  $F(X) = F_{n}(X)$.
%\begin{definition}
\remove{
Let $k =$ argmin $\{ F_{t}(X) = F_{t}+1(X)\} \leq n$. The expansion of $X \subseteq V$ on an influence graph $(G=(E,V),w, f)$, is defined as}
%\end{definition}
%
The \emph{Linear Threshold Rank} of a node $i\in V$ is given by
\[
\LTR(i) = \frac{|F(\{i\} \cup \mathcal{N}(i))|}{n}.
\]

%\medskip
%\noindent
\paragraph{Forward Linear Threshold Rank $-$}
The Forward Linear Threshold Rank~\cite{TFGpau,BGS2021} is a centrality measure very similar to the \LTR, but with a different initial activation set. 
%Given an influence graph $(G=(V,E), w, f)$ of size $n=|V|$,  
Formally, the Forward Linear Threshold Rank of a node $i\in V$ is given by
\[
\FWLTR(i) = \frac{|F(\{i\} \cup \mathcal{N}^+(i))|}{n}
\]
where $\mathcal{N}^+(i) = \{j \in V \mid (i, j) \in E\}$.

\section{Statistical and top measures}
\label{sec:statTopmeasures}

%\bla

%-----------------------------------------------------------------------------------------------------------
%\subsection{Statistical Measures}
\label{sec:statistics}

For analysing the results of a centrality measure on its own, three statistical metrics will be used later in the experimental part: 
\begin{enumerate}
    \item the number of different ranks assigned (\#), 
    \item their standard deviations ($\sigma$) from the mean and,
    \item the Gini coefficient of the ranks. 
\end{enumerate}

The Gini coefficient~\cite{GiniBook,GiniOrigins} comes originally from the field of sociology as a measure of the inequality of populations with respect to different criteria (e.g., wealth spread). The Gini index is often represented graphically through the Lorenz curve~\cite{LorenzCurve}, which shows wealth distribution by plotting the population percentile by income on the horizontal axis and cumulative income on the vertical axis. The Gini coefficient is equal to the area below the line of perfect equality (i.e., $0.5$) minus the area below the Lorenz curve, divided by the area below the line of perfect equality. In other words, it is double the area between the Lorenz curve and the line of perfect equality. 

\begin{definition}
Given a list of values $X$ of size $n$, the Gini coefficient of $X$ is calculated as follows:
$$
\Gini = \frac{\sum_{i=1}^{n} \sum_{j=1}^{n} |x_{i} - x_{j}|}{2n \sum_{i=1}^{n}x_{i}}
$$
\end{definition}

The Gini coefficient is a value in $[0,1]$, where 0 indicates a completely equitable distribution of values, and 1 represents the most radical inequality. This coefficient is lately being very much used in data science as a measure for quantifying the fairness of data distributions coming from other scientific areas.

 \begin{table*}[thp!]
\centering
\caption{Characteristics of the real networks under consideration (in alphabetical order). ACC = Average Clustering Coefficient, MC = size of the main core. When the diameter is $\infty$, the diameter of the biggest connected component is provided.}
\label{tab:networks}
    \centering
    \begin{tabular}{lrrcclcr} 
    \textbf{Data set} & $n$ & $m$& \textbf{Directed?} & \textbf{Edge-weighted?} & \textbf{ACC} & \textbf{Diameter} & \textbf{MC} \\
    \hline
    \href{http://snap.stanford.edu/data/com-Amazon.html}{\Amazon} & 334863 & 925872 & \xmark & \xmark & 0.3967	& 44 & 497\\
    \href{http://snap.stanford.edu/data/ca-GrQc.html}{\ArXiv}                  & 5242     & 14496    & \xmark   & \xmark   & 0.5296  & $\infty$ (17) & 44   \\
    \href{http://snap.stanford.edu/data/as-Caida.html}{\Caida}      & 26475    & 106762   & \cmark  & \cmark  & 0.2082  & 17            & 50   \\
%    \href{http://vlado.fmf.uni-lj.si/pub/networks/data/esna/dining.htm}{Dining-table}           & 26       & 52       & \cmark  & \cmark  & 0.1178  & $\infty$ (6)  & 20   \\
%    \href{http://networkrepository.com/soc-dolphins.php}{Dolphins}               & 62       & 159      & \xmark   & \xmark   & 0.2590  & 8             & 36   \\
     \href{http://snap.stanford.edu/data/email-Enron.html}{\ENRON} & 36692 &	183831 & \xmark & \xmark & 0.4970 & 11 & 275 \\
     \href{http://snap.stanford.edu/data/soc-Epinions1.html}{\Epinions}     & 75879    & 508,837  & \cmark  & \xmark   & 0.1378  & 14            & 422  \\
     \href{http://snap.stanford.edu/data/p2p-Gnutella31.html}{\Gnutella} & 62586 & 147892 & \cmark & \xmark & 0.0055 & 11 & 1004 \\
     \href{http://snap.stanford.edu/data/higgs-twitter.html}{\Higgs}                  & 256491   & 328132   & \cmark  & \cmark  & 0.0156  & 19            & 10   \\
%     \href{http://pitgroup.org/connectome/}{Human brain}  & 480      & 1000     & \xmark   & \cmark  & 0.3004  & $\infty$ (20) & 11   \\
%     \href{http://snap.stanford.edu/data/roadNet-TX.html}{Texas} &	1921660 & 1379917 & \xmark & \xmark & 0.0470 & 1054 & 1579 \\
     \href{http://snap.stanford.edu/data/wiki-Vote.html}{\Wikipedia}   & 7115     & 103689   & \cmark  & \xmark   & 0.1409  & 7             & 336  \\
    %Gowalla      & 196,591  & 950,327  & no   & no   & 0.2367  & 14          & 185                  
    \hline
\end{tabular}
\end{table*}
%-----------------------------------------------------------------------------------------------------------
%\subsection{Top Measures}
\label{sec:tops}

In~\cite{Hasson19} the case of selecting sets of actors with high centrality values was studied
as initial trigger sets. Specifically, they used the Degree, Betweenness and
Eigenvector centralities, but a mixed influence diffusion model was used instead of the usual IC model or LT model, and only two networks of a relatively small size (of the order of a thousand nodes), both of academic collaboration (similar to the \ArXiv network that we use here). However, we were inspired by the idea in that article for the introduction of three
metrics related to the maximization of influence, and we study some variants
that use measures of centrality other than the \FLTR.

The following variables are conceived as natural measures that can allow us to see if the
individual actors with the highest \FLTR in the network are able to exert an amount of influence meaningfully together. 
%\textcolor{red}{Given an influence graph $(G=(V,E), w, f)$ of size $n=|V|$, we define:}
We define:

\paragraph{{\sf Top10} $-$} 
This parameter is the proportion of actors that are influenced if we execute the
influence expansion algorithm taking as initial activating set the 10 actors with the highest \FLTR, i.e.,
\[
\topTEN(G) = \frac{|F(X_{10})|}{n}
\]
where $X_{10}$ is the 10 actors of $G$ with the highest \FLTR. 
% where $X_{10}$ is the set of the 10 actors with the highest value of FLTR. 
%, F (X 10) is the final set of actors that is influenced by taking X 10 as the initial trigger set, and n is the number of vertices of the graph.

\paragraph{{\sf Top10\%Actors} $-$} 
From the \FLTR ranking of actors, let $Y_{10}$ be the 10\% with the greatest capacity to disseminate direct influence. Then,  
\[
\topTENa(G) = \frac{|F(Y_{10})|}{n}
\]

\paragraph{{\sf Top10\%Values} $-$} 
Analogously, let $Z_{10}$ be the set of actors whose influence make up the first 10\% of different ranking values (ties included). Then, 
\[
\topTENv(G) = \frac{|F(Z_{10})|}{n}
\]

\remove{
    \medskip
    We also introduce three variants on the previous metrics, which will characterize, in
    instead of the proportion of nodes that are influenced if we take the actors as the initial set with higher FLTRs, the individual cumulative ability of these to diffuse relative influence to that of all the actors in the network. We do not consider the ability to spread the influence of set itself, but the sum of individual ranks of the actors in the set, divided by the total sum of ranks assigned to all actors in the network. These measures can be used to characterize how the ability to influence is distributed among the individual actors of the network, and if the actors with more centrality really spread much more influence than the rest.
    
    \paragraph{CTop10 $-$} 
    This parameter is the proportion of the total influence spread by all the actors in the network, which corresponds to the top 10 actors in the ranking, i.e., $$\frac{\sum_{i\in X_{10}} FLTR(i)}{\sum_{j\in V} FLTR(j)},$$ where $X_{10}$ is the set of the 10 actors with the highest FLTR and $V$ are all the nodes of the network.
    
    \paragraph{CTop10\%Actors $-$} 
    Analogously to CTop10, we carry out the same calculation but taking in
    instead of the top 10 in the ranking, the top 10\% of actors.
    
    \paragraph{CTop10\%Values $-$} 
    In this case we select the same set described for Top10\%Values,
    but calculating this new measure. We indicate in parentheses the cardinality of the set
    number of actors included in the measure.
}

\section{Networks}
\label{sec:networks}
Table~\ref{tab:networks} summarizes the characteristics of the networks considered for our experiments. The structural characteristics of the networks are described by seven common attributes: the number of vertices, the number of edges, whether the graph is weighted, whether the graph is directed, the average clustering coefficient, and the size of the main core. 

The \textit{average clustering coefficient} ($\mathrm{ACC}$) is the average of the local clustering coefficients in the graph. The local clustering coefficient $C_{i}$ of a node $i$ is the number of triangles $T_{i}$ in which the node participates normalized by the maximum number of triangles that the node could participate in.
\[
\mathrm{ACC} = \frac{1}{n}\sum_{i=1}^{n}C_{i}, \text{\;\;\;\;where\;} 
C_{i} = \frac{T_{i}}{\delta_{i}(\delta_{i}-1)}
\]
where $\delta_{i}$ is the degree of the node $i$, and $n=|V|$. Given a graph $G$ and $k\in\mathbb{Z}^+$, a \emph{$k$-core} is the maximal induced subgraph of $G$ where every node has at least degree $k$. The \emph{main core} is a $k$-core of $G$ with the highest $k$.

\section{Experiments}
\label{sec:experiments}

Although the LT model contemplates an individual threshold for each actor in the network, the existing studies so far have always considered a threshold of 0.5 equal in all actors. What this configuration achieves is implementing a majority activation criterion: a node is activated (or influenced) when at least half of its neighbors are activated.

Our main objective in this work is to begin to study how the dissemination of information on networks behaves when we consider other options for setting those thresholds and how many network actors end up being influenced by this dissemination. Among the experiments performed, we would like to point out here the most interesting ones. The first experiment still considers the same threshold for all the actors but, apart from the standard 0.5, we consider other values in $[0,1]$. The second experiment sets random thresholds to the actors following a random uniform distribution. Finally, the third experiment sets to each node a threshold that depends on its centrality value according to several different measures. All the algorithms for computing the measures were implemented in ANSI C++, using GCC 7.5.0 for compiling the code. The experimental evaluation was performed on a cluster of computers with Intel\textsuperscript{\textregistered} Xeon\textsuperscript{\textregistered} CPU 5670 CPUs of 12 nuclei of 2933 MHz and (in total) 32 Gigabytes of RAM~\cite{rdlab}.

%-----------------------------------------------------------------------------------------------
\subsection{Uniform Thresholds}

The first experiment carried out consists in calculating the \LTR of the network actors in the cases in which the threshold is defined as \nicefrac{1}{4}, \nicefrac{1}{2}, \nicefrac{3}{4} and 1. The results of this experiments are summarized in Table~\ref{tab:uniformThresholds}. For this experiment we obtain six metrics of the distribution of the \FLTR on the actors. 

%Let us remember that the LTR of an actor is the number of actors that he is able to influence, divided by the number of total actors, so we deal with values between 0 and 1. 

% ORIGINAL TABLES:
%\input{TFG-Experiments/Quarterly_FLTR/Amazon}
%\input{TFG-Experiments/Quarterly_FLTR/ArXiv}
%\input{TFG-Experiments/Quarterly_FLTR/Caida}
%\input{TFG-Experiments/Quarterly_FLTR/ENRON}
%\input{TFG-Experiments/Quarterly_FLTR/Epinions}
%\input{TFG-Experiments/Quarterly_FLTR/Gnutella}
%\input{TFG-Experiments/Quarterly_FLTR/Higgs}
%\input{TFG-Experiments/Quarterly_FLTR/Texas}
%\input{TFG-Experiments/Quarterly_FLTR/Wikipedia}

\begin{table*}[thp!]
\centering
\caption{Experimental results when considering uniform thresholds for $\theta\in\{\nicefrac{1}{4}, \nicefrac{1}{2}, \nicefrac{3}{4}, 1.00\}$.}
\label{tab:uniformThresholds}
    \centering
\scalebox{0.9}{\vbox{\begin{tabular}{lrrrlllrlr}
    \cline{2-10}
    &
    \textbf{$\theta$} & 
    \multicolumn{1}{c}{\textbf{$\sigma$}} & 
    \multicolumn{1}{c}{\textbf{\#}} & 
    \multicolumn{1}{c}{\textbf{Gini}} & 
    \multicolumn{1}{c}{\textbf{\topTEN}} & 
    \multicolumn{2}{c}{\textbf{\topTENa}} & 
    \multicolumn{2}{c}{\textbf{\topTENv}} \\
%    \hline
%    \multicolumn{9}{c}{\textbf{Amazon}}\\
    \cline{2-10}
    \parbox[t]{2mm}{\multirow{4}{*}{\rotatebox[origin=c]{90}{\textbf{\Amazon}}}} 
    & 0.25 &  0.001358 &  1146 &  0.697121 &   0.00658478 &   0.754688 & (33486) &    0.628117 & (518) \\
    & 0.50 &  0.000037 &   224 &  0.411186 &  0.000979505 &   0.301711 & = &   0.00142745 & (25) \\
    &  0.75 &  0.000031 &   203 &  0.372209 &   0.00078241 &   0.210316 & = &  0.000958601 & (24) \\
    & 1.00 &  0.000017 &   153 &  0.326807 &   2.9863e-05 &  0.0999991 & = &   5.0767e-05 & (17) \\
%    \hline
%    \multicolumn{9}{c}{\textbf{ArXiv}}\\
    \cline{2-10}
    \parbox[t]{2mm}{\multirow{4}{*}{\rotatebox[origin=c]{90}{\textbf{\ArXiv}}}} 
    & 0.25 &  0.149279 &  435 &  0.854073 &    0.648417 &   0.670355 & (524) &   0.656429 & (104) \\
    & 0.50 &  0.005950 &  169 &  0.540466 &  0.00267074 &   0.263068 & = &  0.00839374 & (20) \\
    & 0.75 &  0.004339 &  137 &  0.519565 &  0.00209844 &   0.167684 & = &   0.0034338 & (14) \\
    & 1.00 &  0.003578 &  118 &  0.513171 &  0.00190767 &  0.0999618 & = &   0.0022892 & (12) \\
    %\hline
    %\multicolumn{9}{c}{\textbf{Caida}}\\
    \cline{2-10}
    \parbox[t]{2mm}{\multirow{4}{*}{\rotatebox[origin=c]{90}{\textbf{\Caida}}}} 
    & 0.25 &  0.234984 &  2033 &  0.827517 &     0.890954 &   0.979339 & (2647) &   0.979339 & (2426) \\
    & 0.50 &  0.014930 &  1234 &  0.704623 &     0.114145 &   0.555694 & = &    0.362606 & (137) \\
    & 0.75 &  0.012896 &  1174 &  0.693649 &    0.0957129 &    0.47169 & = &    0.266931 & (130) \\
    & 1.00 &  0.001261 &   158 &  0.503259 &  0.000377715 &  0.0999811 & = &  0.000642115 & (17) \\
    %\hline
    %\multicolumn{9}{c}{\textbf{ENRON}}\\
    \cline{2-10}
    \parbox[t]{2mm}{\multirow{4}{*}{\rotatebox[origin=c]{90}{\textbf{\ENRON}}}} 
    & 0.25 &  0.268885 &  2786 &  0.833348 &     0.110406 &   0.845307 & (3669) &    0.841655 & (426) \\
    & 0.50 &  0.034267 &  2322 &  0.752884 &    0.0292707 &   0.588139 & = &    0.442222 & (236) \\
    & 0.75 &  0.013211 &  1852 &  0.715277 &    0.0168429 &    0.39202 & = &    0.148643 & (190) \\
    & 1.00 &  0.009538 &  1448 &  0.726360 &  0.000272539 &  0.0999945 & = &  0.00411534 & (151) \\
    %\hline
    %\multicolumn{9}{c}{\textbf{Epinions}}\\
    \cline{2-10}
    \parbox[t]{2mm}{\multirow{4}{*}{\rotatebox[origin=c]{90}{\textbf{\Epinions}}}} 
    & 0.25 &  0.034836 &  1395 &  0.961041 &    0.0462051 &   0.612976 & (7587) &    0.609418 & (375) \\
    & 0.50 &  0.001259 &   828 &  0.830114 &     0.014879 &   0.484587 & = &    0.0422778 & (85) \\
    & 0.75 &  0.001185 &   814 &  0.828013 &    0.0144045 &   0.370326 & = &    0.0377838 & (84) \\
    & 1.00 &  0.000343 &   327 &  0.707198 &  0.000131789 &  0.0999881 & = &  0.000487618 & (37) \\
    %\hline
    %\multicolumn{9}{c}{\textbf{Gnutella}}\\
    \cline{2-10}
    \parbox[t]{2mm}{\multirow{4}{*}{\rotatebox[origin=c]{90}{\textbf{\Gnutella}}}} 
    & 0.25 &  0.299963 &  593 &  0.892894 &    0.974148 &   0.979692 & (6258) &  0.979788 & (6716) \\
    & 0.50 &  0.000173 &  111 &  0.689648 &  0.00615154 &    0.97402 & = &  0.00672674 & (13) \\
    & 0.75 &  0.000172 &  111 &  0.689222 &  0.00594382 &   0.956076 & = &  0.00640718 & (12) \\
    & 1.00 &  0.000070 &   52 &  0.545161 &  0.00015978 &  0.0999904 & = &  0.000111846 & (7) \\
    %\hline
    %\multicolumn{9}{c}{\textbf{Higgs}}\\
    \cline{2-10}
    \parbox[t]{2mm}{\multirow{4}{*}{\rotatebox[origin=c]{90}{\textbf{\Higgs}}}} 
    & 0.25 &  0.000010 &  77 &  0.279112 &  0.000557524 &   0.147467 & (25649) &  0.000557524 & (10) \\
    & 0.50 &  0.000006 &  52 &  0.218244 &   0.00035089 &    0.13094 & = &   0.000323598 & (8) \\
    & 0.75 &  0.000006 &  53 &  0.214459 &  0.000362586 &   0.125685 & = &   0.000304104 & (6) \\
    & 1.00 &  0.000005 &  46 &  0.192700 &  3.89877e-05 &  0.0999996 & = &   2.33926e-05 & (6) \\
    %\hline
    %\multicolumn{9}{c}{\textbf{Texas}}\\
%    \cline{2-10}
%    \parbox[t]{2mm}{\multirow{4}{*}{\rotatebox[origin=c]{90}{\textbf{\Texas}}}} 
%    & 0.25 &  0.424542 &  3226 &  0.361401 &     0.907126 &   0.966586 & (137991) &  0.92991 & (27780)  \\
%    & 0.50 &  0.000003 &    35 &  0.200820 &  7.24681e-06 &   0.128752 & = &  4.34809e-06 & (6) \\
%    & 0.75 &  0.000002 &    28 &  0.202959 &  7.24681e-06 &   0.111038 & = &  2.17404e-06& (3) \\
%    & 1.00 &  0.000002 &    26 &  0.184078 &  7.24681e-06 &  0.0999995 & = &  2.17404e-06 & (3) \\
    %\hline
    %\multicolumn{9}{c}{\textbf{Wikipedia}}\\
    \cline{2-10}
    \parbox[t]{2mm}{\multirow{4}{*}{\rotatebox[origin=c]{90}{\textbf{\Wikipedia}}}} 
    & 0.25 &  0.026782 &  309 &  0.875415 &     0.32298 &   0.366268 & (711) &    0.324666 & (37) \\
    & 0.50 &  0.006443 &  234 &  0.766315 &  0.00196767 &   0.359663 & = &  0.00477864 & (25) \\
    & 0.75 &  0.006125 &  238 &  0.763096 &  0.00196767 &   0.288686 & = &  0.00463809 & (26) \\
    & 1.00 &  0.005943 &  238 &  0.761196 &  0.00140548 &  0.0999297 & = &  0.00337316 & (24) \\
    \cline{2-10}
    \end{tabular}}}
\end{table*}

Observing the sharp differences obtained in the results of this first experiment, we decided to repeat it for the \Gnutella and \ArXiv networks for values of the threshold between 0.2 and 0.5, and for \Amazon, \Caida, \ENRON, \Epinions and \Wikipedia between 0.2 and 0.4. On the contrast, the remaining networks showed a considerable degree of monotony in metrics despite threshold variation.  Table~\ref{tab:uniformThresholds2} shows the refined results for the \Caida and \Gnutella networks, while Table~\ref{tab:uniformThresholds3} in the Appendix contains the results for the remaining ones. Here we only expose and comment on the most relevant results.

The extreme variation of \topTENv due to widespread ties in the allocation of the
\FLTR questions its usefulness as a metric for maximizing influence, since its
behavior is not justified by the variation in the other measures.

Note that there is an obvious correlation between \topTENa and the \gini, since the more
unequal is a distribution, the greater proportion of the value will be accumulated by a small group of actors. However, this does not necessarily occur in \topTEN, where the set of actors The observed size has a fixed size of 10, regardless of the size of the network considered. We show the value of the \gini to decrease when the influence threshold is increased, but \topTEN remains behavior totally independent of this trend, as we clearly see in \Wikipedia or \Caida (see Table~\ref{tab:uniformThresholds2}).

We see a curious rise and fall of the cardinality of the set of \topTENv
between the thresholds of 0.26 and 0.34 in the \Gnutella network  (see Table~\ref{tab:uniformThresholds2}), which is not reflected in the variation of the
other metrics for this range of thresholds. There is also a sudden drop from
0.48, reflected in all metrics except \topTENa, which suffers a relatively decrease
small. This suggests that despite having been radically affected the distribution of
\FLTR of the actors and to have severely limited the influence of the influence suddenly, in the networks observed 10\% of the most influential actors is a set large enough and
robust enough to continue to expand its influence to most of the \Gnutella network
at this threshold magnitude. This behavior is not observed in the other social network that has been considered between 0.2 and 0.5, \ArXiv, where \topTENa follow an online behavior with the other metrics. Despite \Gnutella's dimensions being larger than \ArXiv's,
the most significant differences between these two networks, in terms of the parameters that we have obtained from them, is that \Gnutella is directed and \ArXiv is not, and \ArXiv has a diameter and ACC more high, and a much smaller Main Core. In the case of ACC, it is of the order of 100 times greater. These data suggest that \ArXiv is a much more interconnected and distributed network than \Gnutella, whose parameters indicate that most nodes have few interconnections, but there are a subset of nodes with very high degrees, which would be very capable of diffusing the influence effectively, and therefore would be in the \topTENa. \Gnutella's \gini is especially high compared to these networks, a fact that could validate this explanation.

The results obtained up to this point confirm that in most of the treated networks, the
range of threshold values of the resistance to be influenced more relevant to the maximization of the influence is approximately at $[0.2, 0.5]$, where it is possible to accumulate greater proportion of the total influence diffused in groups of actors of small size.

\begin{table*}[thp!]
\centering
\caption{Experimental results when considering uniform thresholds $\theta$: refined zoom in $[0.2,~0.4]$ and $[0.2,~0.5]$.}
\label{tab:uniformThresholds2}
    \centering
\scalebox{0.9}{\vbox{\begin{tabular}{lrrrlllrlr}
    \cline{2-10}
    &
    \textbf{$\theta$} & 
    \multicolumn{1}{c}{\textbf{$\sigma$}} & 
    \multicolumn{1}{c}{\textbf{\#}} & 
    \multicolumn{1}{c}{\textbf{Gini}} & 
    \multicolumn{1}{c}{\textbf{\topTEN}} & 
    \multicolumn{2}{c}{\textbf{\topTENa}} & 
    \multicolumn{2}{c}{\textbf{\topTENv}} \\
    %\hline
    %\multicolumn{9}{c}{\textbf{Caida}}\\
    \cline{2-10}
    \parbox[t]{2mm}{\multirow{11}{*}{\rotatebox[origin=c]{90}{\textbf{\Caida}}}} 
  & 0.20 &  0.365273 &  1455 &  0.747064 &  0.907347 &  0.995392 & (2647) &  0.995392 & (6562) \\
  & 0.22 &  0.338180 &  1615 &  0.775445 &  0.906327 &  0.995392 & = &  0.995392 & (5223) \\
  & 0.24 &  0.322551 &  1740 &  0.790862 &  0.904363 &   0.98916 & = &  0.994334 & (4594) \\
  & 0.26 &  0.218601 &  2077 &  0.825349 &  0.871124 &  0.978357 & = &  0.978357 & (2225) \\
  & 0.28 &  0.197560 &  2173 &  0.822601 &  0.860057 &  0.933975 & = &  0.930727 & (1818) \\
  & 0.30 &  0.142933 &  2291 &  0.792907 &  0.787762 &   0.92729 & = &  0.918829 & (1399) \\
  & 0.32 &  0.130483 &  2355 &  0.783452 &  0.765703 &  0.913957 & = &  0.899452 & (1302) \\
  & 0.34 &  0.062324 &  2366 &  0.720697 &  0.187875 &   0.87101 & = &   0.795958 & (408) \\
  & 0.36 &  0.057867 &  2395 &  0.714390 &   0.18085 &  0.859754 & = &   0.791917 & (386) \\
  & 0.38 &  0.049777 &  2424 &  0.704256 &  0.184816 &  0.849745 & = &   0.769481 & (334) \\
  & 0.40 &  0.044526 &  2402 &  0.697489 &  0.171709 &  0.838414 & = &   0.745647 & (277) \\
    %\hline
    %\multicolumn{9}{c}{\textbf{Gnutella}}\\
    \cline{2-10}
    \parbox[t]{2mm}{\multirow{10}{*}{\rotatebox[origin=c]{90}{\textbf{\Gnutella}}}} 
    & 0.20,~0.22 &  0.373341 &  280 &  0.819895 &    0.974547 &  0.988176 & (6258) &   0.987873 & (212) \\
%    & 0.22 &  0.373341 &  281 &  0.819895 &    0.974547 &  0.988176 & = &   0.987873 & = \\
    & 0.24 &  0.373303 &  277 &  0.819943 &    0.974547 &  0.988176 & = &   0.987361 & (186) \\
    & 0.26 &  0.299963 &  593 &  0.892894 &    0.974148 &  0.979692 & = &  0.979788 & (6716) \\
    & 0.28 &  0.299944 &  594 &  0.892910 &    0.974148 &  0.979692 & = &  0.979788 & (6715) \\
    & 0.30,~0.32 &  0.297771 &  615 &  0.894629 &    0.974148 &  0.979596 & = &  0.979596 & (6606) \\
%    & 0.32 &  0.297771 &  615 &  0.894629 &    0.974148 &  0.979596 & = &  0.979596 & = \\
    & 0.34,~0.38 &  0.090475 &  581 &  0.986210 &    0.971863 &  0.981753 & = &   0.971863 & (610) \\
%    & 0.36 &  0.090475 &  581 &  0.986210 &    0.971863 &  0.981753 & = &   0.971863 & = \\
%    & 0.38 &  0.090475 &  581 &  0.986211 &    0.971863 &  0.981721 & = &   0.971863 & = \\
    & 0.40,~0.42 &  0.080219 &  666 &  0.986708 &    0.971783 &  0.981657 & = &   0.971783 & (495) \\
%    & 0.42 &  0.080219 &  666 &  0.986708 &    0.971783 &  0.981657 & = &   0.971783 & = \\
    & 0.44 &  0.080126 &  663 &  0.986709 &    0.971783 &  0.981657 & = &   0.971783 & (497) \\
    &  0.46,~0.48 &  0.080126 &  662 &  0.986709 &    0.971783 &  0.981657 & = &   0.971783 & (495) \\
%    &  0.48 &  0.080126 &  662 &  0.986709 &    0.971783 &  0.981657 & = &   0.971783 & = \\
    &  0.50 &  0.000173 &  111 &  0.689648 &  0.00615154 &   0.97402 & = &  0.00672674 & (13) \\
    \cline{2-10}
    \end{tabular}}}
\end{table*}

\begin{table*}[thp!]
\centering
\caption{Experimental results (for 100 executions) for random thresholds under a random uniform distribution.}
\label{tab:randomThresholds1}
    \centering
\scalebox{0.9}{\vbox{\begin{tabular}{lrrrlllrlr}
    \hline
    \textbf{Network} &
    \multicolumn{1}{c}{\textbf{$\sigma$}} & 
    \multicolumn{1}{c}{\textbf{\#}} & 
    \multicolumn{1}{c}{\textbf{Gini}} & 
    \multicolumn{1}{c}{\textbf{\topTEN}} & 
    \multicolumn{2}{c}{\textbf{\topTENa}} & 
    \multicolumn{2}{c}{\textbf{\topTENv}} \\
    \hline
    \Amazon &  0.007479 &   2317 &  0.002218 &           1 &         1 & (33486) &          1 & (1492) \\
    \ArXiv &  0.316806 &   1478 &  0.208256 &    0.793209 &    0.793209 & (524) &    0.793209 & (234) \\
    \Caida &  0.007981 &   3248 &  0.002557 &           1 &          1 & (2647) &           1 & (396) \\
    \ENRON &  0.248287 &  13117 &  0.083801 &    0.918347 &   0.918347 & (3669) &   0.918347 & (1390) \\
    \Epinions &  0.269755 &   5816 &  0.259223 &    0.628316 &   0.637106 & (7587) &    0.628316 & (636) \\
    \Gnutella &  0.405467 &   2288 &  0.768062 &    0.972805 &   0.992394 & (6258) &    0.978238 & (264) \\
    \Higgs &  0.003500 &   9404 &  0.709183 &  0.00924399 &  0.139252 & (25649) &  0.0245662 & (1472) \\
    \Wikipedia &  0.142819 &   1048 &  0.277731 &    0.327056 &    0.383696 & (711) &    0.333942 & (109) \\
    \hline
    \end{tabular}}}
\end{table*}
%-----------------------------------------------------------------------------------------------
\subsection{Random Thresholds}

In this experiment, we set random thresholds to the actors according to a random uniform distribution in $[0,1]$, and later on two different intervals.

\subsubsection{Random Uniform Distribution}

For this experiment we assigned thresholds in $[0,1]$ to the actors
of the network following a uniform probability distribution. To determine centrality
of the actors we execute a certain number of instances of the expansion algorithm of
influence, each with new randomly assigned values, and finally we get the
average \FLTR of the actors in these executions. Table~\ref{tab:randomThresholds1} shows the results for this experiment for 100 executions. 

%Since we have observed that the differences between 100 runs and 20 are very small, and the difference in run time of the experiments is substantial, we chose to run the algorithm 20 times and extract the mean of the \FLTR obtained. 

%The Running this experiment took approximately 90 CPU hours in total.

We can see that there are no large differences for any of the variables between the
two experiments, and that for the vast majority of networks, a uniform random distribution is very conducive to maximizing influence, achieving very satisfactory results
with only the top 10 actors in the ranking. However, we also see that rarely
we achieve more expansion of influence if we increase the set to the first 10% of actors
or securities. We also see for the first time, that the \topTEN sets have managed to expand
the influence of all the actors in the Amazon and \Caida networks.
\begin{table*}[thp!]
\centering
\caption{Experimental results (for 20 executions) for random thresholds under a random uniform distribution within $(0,0.5]$.}
\label{tab:randomThresholds2}
    \centering
\scalebox{0.9}{\vbox{\begin{tabular}{lrrrlllrlr}
    \hline
    \textbf{Network} &
    \multicolumn{1}{c}{\textbf{$\sigma$}} & 
    \multicolumn{1}{c}{\textbf{\#}} & 
    \multicolumn{1}{c}{\textbf{Gini}} & 
    \multicolumn{1}{c}{\textbf{\topTEN}} & 
    \multicolumn{2}{c}{\textbf{\topTENa}} & 
    \multicolumn{2}{c}{\textbf{\topTENv}} \\
    \hline
       \Amazon &  0.021248 &   519 &  0.008737 &           1 &         1 & (33486) &          1 & (1192) \\
        \ArXiv &  0.316359 &   646 &  0.213341 &    0.793209 &    0.793209 & (524) &    0.793209 & (151) \\
        \Caida &  0.031134 &   526 &  0.015298 &           1 &          1 & (2647) &           1 & (117) \\
        \ENRON &  0.249355 &  3265 &  0.089727 &    0.918347 &   0.918347 & (3669) &    0.918347 & (380) \\
     \Epinions &  0.266832 &   702 &  0.269977 &    0.629397 &     0.7452 & (7587) &   0.763597 & (9994) \\
     \Gnutella &  0.407975 &   424 &  0.768950 &    0.974579 &   0.991276 & (6258) &  0.991628 & (12851) \\
        \Higgs &  0.003177 &  4318 &  0.714451 &  0.00971964 &  0.139479 & (25649) &  0.0190182 & (1027) \\
    \Wikipedia &  0.139828 &   571 &  0.285285 &    0.327056 &    0.374982 & (711) &     0.334505 & (77) \\
    \hline
    \end{tabular}}}
%\end{table*}
%
\bigskip
%
%\begin{table*}[p!]
\centering
\caption{Experimental results (for 100 executions) for random thresholds under a random uniform distribution within $[0.5,1]$.}
\label{tab:randomThresholds3}
    \centering
\scalebox{0.9}{\vbox{    \begin{tabular}{lrrrlllrlr}
    \hline
    \textbf{Network} &
    \multicolumn{1}{c}{\textbf{$\sigma$}} & 
    \multicolumn{1}{c}{\textbf{\#}} & 
    \multicolumn{1}{c}{\textbf{Gini}} & 
    \multicolumn{1}{c}{\textbf{\topTEN}} & 
    \multicolumn{2}{c}{\textbf{\topTENa}} & 
    \multicolumn{2}{c}{\textbf{\topTENv}} \\
    \hline
       \Amazon &  0.163244 &  35500 &  0.712485 &    0.717867 &  0.719664 & (33486) &   0.717935 & (9784) \\
        \ArXiv &  0.228914 &   2007 &  0.240972 &    0.791683 &    0.791683 & (524) &    0.791683 & (240) \\
        \Caida &  0.246121 &   7448 &  0.609839 &    0.926308 &   0.926308 & (2647) &    0.926308 & (818) \\
        \ENRON &  0.208083 &  13455 &  0.152003 &    0.918075 &   0.918075 & (3669) &   0.918075 & (1572) \\
     \Epinions &  0.124726 &  11641 &  0.759954 &    0.537092 &   0.542825 & (7587) &   0.537092 & (1365) \\
     \Gnutella &  0.000176 &   1059 &  0.691910 &  0.00560828 &   0.973988 & (6258) &    0.856725 & (145) \\
        \Higgs &  0.000043 &   2344 &  0.528186 &    0.002234 &  0.141759 & (25649) &  0.00643687 & (358) \\
    \Wikipedia &  0.091907 &   1960 &  0.678521 &    0.319606 &    0.364301 & (711) &    0.331553 & (214) \\
    \hline
    \end{tabular}}}
\end{table*}

We observe that the number of different values produced in the ranking is much higher
in all cases to those of the \FLTR with the threshold equal to 0.5 (see Table~\ref{tab:uniformThresholds}) and are more dispersed, as evidenced by the standard deviation $\sigma$. The \gini also indicates that in general the distribution of ranking values is significantly more equal than when the threshold is fixed at 0.5. We see a significant reduction in \gini in almost all networks (e.g., in the case of \ENRON, it drops approximately from 0.75 to 0.08). The exceptions are in \Gnutella and \Higgs, where there is an ascent with respect to the fixed threshold.

\subsubsection{Random Uniform Distribution in Two Different Intervals}

As a result of the previous experiment, we wondered what would happen if we assigned the thresholds of uniform random way for values in (0, 0.5] and in [0.5, 1]. 
By selecting these two intervals we can see what happens when the
thresholds are in the more permissive or restrictive half for the transmission of influence.
In Table~\ref{tab:randomThresholds2} and Table~\ref{tab:randomThresholds3} we can check the results.

%The duration of this experiment was approximately 24 hours. 

%\bigskip
%

%-----------------------------------------------------------------------------------------------

The values of the variable \topTEN in Table~\ref{tab:randomThresholds3} suggest 
that for most of networks it is possible to achieve a good expansion of influence even when the thresholds of resistance are distributed in the most restrictive half of the interval. In some cases, we obtain practically the same expansion as when assigning the thresholds in $(0, 0.5]$, as is the case for the networks \ArXiv, \Wikipedia or \ENRON. As expected, a smaller value is reached in general. The values of \# are notably different between the two experiments, being the values of the second significantly larger than those of the first in almost all cases. This may indicate that, when the thresholds tend to be more restrictive, there is a more diverse ranking. Despite this, there does not appear to be any trend in the value of the \gini between the two experiments, and the standard deviation increases for some networks and decreases for others.

If we compare Table~\ref{tab:randomThresholds2} with Table~\ref{tab:randomThresholds1} in the previous section, we observe little difference in the variables that measure the expansion of influence. In fact, the only relevant improvement occurs on the \Epinions network. We also see a general reduction in the number of rankings assigned to the actors (\#), but very similar standard deviation and \gini values. These results indicate that a random distribution of the thresholds behaves practically the same with respect to the diffusion of influence whether they are in $(0, 0.5]$ or $(0, 1]$, suggesting that perhaps the fact that a minority fraction of the actors have high thresholds affects little to the ability to influence of the actors in general.

\subsection{Thresholds Fixed by Other Centrality Measures}

In this experiment we will use different centrality criteria to set the thresholds of the actors of a network. We will denote \FLTR$_{\mu}$ to the forward linear threshold rank of the actors in a social network, where its threshold of resistance to influence is equal to its centrality calculated according to method $\mu$, for example \FLTR$_{ICR}$ is the \FLTR ranking of the actors in a network when we set the resistance threshold of each actor to its \ICR centrality value. 
We introduce these metrics to study the possibility that an actor's resistance to being influenced in a network is related to its centrality value, perhaps by a topological criterion or by own ability to spread influence, and in turn, we study the case that this related with the lack of these qualities. In Table~\ref{tab:randomThresholds4} at the Appendix, we can see how the thresholds are assigned according to the different centrality measures in each network. 

% TABLES

\begin{table*}[thp!]
\centering
%\caption{Experimental results when the thresholds of the nodes are assigned according to their value of Betweenness measure.}
\caption{Experimental results for node thresholds assigned according to \Betweenness.}
\label{tab:randomThresholds5}
    \centering
    \scalebox{0.9}{\vbox{
    \begin{tabular}{lrrrlllrlr}
    \hline
    \textbf{Network} &
    \multicolumn{1}{c}{\textbf{$\sigma$}} & 
    \multicolumn{1}{c}{\textbf{\#}} & 
    \multicolumn{1}{c}{\textbf{Gini}} & 
    \multicolumn{1}{c}{\textbf{\topTEN}} & 
    \multicolumn{2}{c}{\textbf{\topTENa}} & 
    \multicolumn{2}{c}{\textbf{\topTENv}} \\
    \hline
       \Amazon &  0.144358 &  185 &  0.021906 &    0.986335 &  0.996363 & (33486) &   0.987983 & (572) \\
        \ArXiv &  0.321546 &   84 &  0.207490 &    0.793209 &    0.793209 & (524) &    0.793209 & (11) \\
        \Caida &  0.097087 &   10 &  0.009517 &           1 &          1 & (2647) &        1 & (26228) \\
        \ENRON &  0.269095 &  431 &  0.104976 &    0.915458 &   0.918347 & (3669) &  0.915458 & (1392) \\
     \Epinions &  0.277439 &   39 &  0.265431 &    0.628725 &   0.757126 & (7587) &    0.629173 & (28) \\
     \Gnutella &  0.410393 &   73 &  0.767687 &    0.974978 &   0.996421 & (6258) &     0.974579 & (9) \\
        \Higgs &  0.003131 &  198 &  0.817138 &  0.00899057 &  0.148863 & (25649) &  0.0103551 & (109) \\
    \Wikipedia &  0.145380 &   14 &  0.275292 &     0.32818 &    0.431061 & (711) &    0.339002 & (55) \\
    \hline
    \end{tabular}
    }}
%\end{table*}
%
\bigskip
%
%\begin{table*}[hp!]
\centering
%\caption{Experimental results when the thresholds of the nodes are assigned according to their value of ICR.}
\caption{Experimental results for node thresholds assigned according to their \ICR value.}
\label{tab:randomThresholds6}
    \centering
    \scalebox{0.9}{\vbox{
    \begin{tabular}{lrrrlllrlr}
    \hline
    \textbf{Network} &
    \multicolumn{1}{c}{\textbf{$\sigma$}} & 
    \multicolumn{1}{c}{\textbf{\#}} & 
    \multicolumn{1}{c}{\textbf{Gini}} & 
    \multicolumn{1}{c}{\textbf{\topTEN}} & 
    \multicolumn{2}{c}{\textbf{\topTENa}} & 
    \multicolumn{2}{c}{\textbf{\topTENv}} \\
    \hline
       \Amazon &  0.048237 &   159 &  0.002346 &    0.997814 &        1 & (33486) &   0.998719 & (67) \\
        \ArXiv &  0.016946 &   389 &  0.626108 &   0.0721099 &   0.603396 & (524) &   0.183518 & (43) \\
        \Caida &  0.058035 &  3025 &  0.664515 &    0.355354 &   0.95966 & (2647) &  0.913692 & (364) \\
        \ENRON &  0.016987 &  2322 &  0.691726 &   0.0317235 &  0.526927 & (3669) &  0.251199 & (237) \\
     \Epinions &  0.003255 &  1614 &  0.862370 &   0.0452035 &  0.599573 & (7587) &  0.152493 & (169) \\
     \Gnutella &  0.387287 &   267 &  0.801691 &    0.974643 &  0.990701 & (6258) &  0.983383 & (106) \\
        \Higgs &  0.000028 &   144 &  0.448542 &  0.00132948 &  0.15022 & (25649) &  0.0015907 & (18) \\
    \Wikipedia &  0.028498 &   781 &  0.744640 &    0.113563 &   0.355727 & (711) &   0.195643 & (92) \\
    \hline
    \end{tabular}
    }}
%\end{table*}
%
\bigskip
%
%\begin{table*}[hp!]
\centering
%\caption{Experimental results when the thresholds of the nodes are assigned according to their value of PageRank.}
\caption{Experimental results for node  thresholds assigned according to \PageRank.}
\label{tab:randomThresholds7}
    \centering
    \scalebox{0.9}{\vbox{
    \begin{tabular}{lrrrlllrlr}
    \hline
    \textbf{Network} &
    \multicolumn{1}{c}{\textbf{$\sigma$}} & 
    \multicolumn{1}{c}{\textbf{\#}} & 
    \multicolumn{1}{c}{\textbf{Gini}} & 
    \multicolumn{1}{c}{\textbf{\topTEN}} & 
    \multicolumn{2}{c}{\textbf{\topTENa}} & 
    \multicolumn{2}{c}{\textbf{\topTENv}} \\
    \hline
       \Amazon &  0.000000 &    1 &  0.000000 &           1 &         1 & (33486) &      1 & (334863) \\
        \ArXiv &  0.321327 &   84 &  0.207110 &    0.793209 &    0.793209 & (524) &   0.793209 & (11) \\
        \Caida &  0.066887 &    7 &  0.004494 &           1 &          1 & (2647) &       1 & (26356) \\
        \ENRON &  0.256195 &  360 &  0.085229 &    0.918347 &   0.918347 & (3669) &   0.918347 & (37) \\
     \Epinions &  0.275013 &   39 &  0.258302 &    0.628725 &   0.756837 & (7587) &   0.629173 & (28) \\
     \Gnutella &  0.410393 &   73 &  0.767687 &    0.974978 &   0.996421 & (6258) &    0.974579 & (9) \\
        \Higgs &  0.003319 &  189 &  0.785266 &  0.00894768 &  0.151744 & (25649) &  0.0101797 & (83) \\
    \Wikipedia &  0.145312 &   14 &  0.274871 &     0.32832 &    0.431061 & (711) &   0.339002 & (55) \\
    \hline
    \end{tabular}
    }}
%\end{table*}
%
\bigskip
%
%\begin{table*}[hp!]
\centering
\caption{Experimental results when the thresholds of the nodes are assigned according to their value of \FLTR.}
\label{tab:randomThresholds8}
    \centering
    \scalebox{0.9}{\vbox{
    \begin{tabular}{lrrrlllrlr}
    \hline
    \textbf{Network} &
    \multicolumn{1}{c}{\textbf{$\sigma$}} & 
    \multicolumn{1}{c}{\textbf{\#}} & 
    \multicolumn{1}{c}{\textbf{Gini}} & 
    \multicolumn{1}{c}{\textbf{\topTEN}} & 
    \multicolumn{2}{c}{\textbf{\topTENa}} & 
    \multicolumn{2}{c}{\textbf{\topTENv}} \\
    \hline
       \Amazon &  0.004888 &    4 &  0.000024 &           1 &         1 & (33486) &       1 & (334855) \\
        \ArXiv &  0.322199 &   84 &  0.208631 &    0.793209 &    0.793209 & (524) &    0.793209 & (11) \\
        \Caida &  0.151072 &   20 &  0.023374 &           1 &          1 & (2647) &        1 & (25895) \\
        \ENRON &  0.296371 &  385 &  0.118261 &    0.918347 &   0.918347 & (3669) &    0.918347 & (39) \\
     \Epinions &  0.277610 &   39 &  0.265945 &    0.628725 &   0.757087 & (7587) &    0.629173 & (28) \\
     \Gnutella &  0.410393 &   73 &  0.767687 &    0.974978 &   0.996421 & (6258) &     0.974579 & (9) \\
        \Higgs &  0.003712 &  183 &  0.706886 &  0.00899837 &  0.157916 & (25649) &  0.0103473 & (101) \\
    \Wikipedia &  0.146984 &   14 &  0.285386 &     0.32832 &    0.430921 & (711) &    0.338721 & (54) \\
    \hline
    \end{tabular}}}
\end{table*}

\smallskip
\noindent
%\paragraph
\textbf{FLTR$_{\Betweenness}$ $-$} 
When fixing the threshold according to \Betweenness (see Table~\ref{tab:randomThresholds5}), we can observe similar results to those of the experiment detailed in
Table~\ref{tab:randomThresholds1} for the thresholds distributed in a uniform random way. The standard deviation and the \gini are also very similar. Where the results of these two 
experiments really differ is in the number of distinct values in the ranking, which are significantly smaller now. These results suggest that we can expect achieve a very good expansion with very small initial sets, but note that we gain little or nothing by increasing the pool of the top 10 to the first 10\%.

\smallskip
\noindent
%\paragraph
\textbf{FLTR$_{\ICR}$ $-$} 
When fixing the threshold according to \ICR (see Table~\ref{tab:randomThresholds6}) the values of \topTEN, \topTENa and \topTENv are lower than when fixed according to \Betweenness (compare with Table~\ref{tab:randomThresholds5}) for all networks except \Amazon and \Higgs, where we see a slight increase. The values of $\sigma$ are also lower, indicating higher concentration, and gives more variety of values, which are also more unequally distributed, as indicated the \gini, except in the \Higgs network, which undergoes the opposite behavior.

\smallskip
\noindent
%\paragraph
\textbf{FLTR$_{\PageRank}$ and FLTR$_{\FLTR}$ $-$} 
When fixing the threshold according to \PageRank and \FLTR (see Tables~\ref{tab:randomThresholds7} and~\ref{tab:randomThresholds8}, respectively) present extremely similar values in the variables related to the expansion of influence. We could expect similarities given the values in Table~\ref{tab:randomThresholds4}, where we observe that they give very comparable threshold assignments, however, it is surprising to what extent the diffusion of the influence on the \topTEN variables between both methods. The values obtained for the metrics $\sigma$, \# and Gini are not disparate, which show similar results under both centralities. From these tables we deduce that assigning the thresholds according to the \Pagerank and \FLTR centralities of the actors generates favorable conditions for a great influence expansion from reduced sets. In nearly all cases, it is enough to have the first 10 actors of the ranking to reach the entire network or its vast majority, except for two networks (\Higgs and \Wikipedia) whose individual characteristics seem to imply difficulties for the diffusion of influence, as by this and previous experiments suggest.

\subsubsection{Complementaries}

We have also performed an additional experiment that complements the previous one. We will denote by \FLTR$_{1-\mu}$ the case in which the thresholds are set by the complement of the value obtained from the centrality $\mu$, that is, if an actor obtains a centrality value $c$ in the measure $\mu$, we will assign a threshold equivalent to $1-c$. In this way we can observe the case in which a greater centrality corresponds to a greater resistance to influence, and the opposite.

The \FLTR$_{1-\Betweenness}$ (see Table~\ref{tab:randomThresholds55} in the Appendix) behaves in the opposite way in the \topTEN variables, where we barely managed to expand the influence. The values of the \gini they have also been reversed: the high values in Table~\ref{tab:randomThresholds5} are here low, and vice versa. The standard deviation $\sigma$ is very low in all cases, indicating that ranking values are produced very clustered around the mean.

In the \FLTR$_{1-\ICR}$ (see Table~\ref{tab:randomThresholds66} in the Appendix) we see that we hardly managed to diffuse the influence in none of the \topTEN variables. We also have low values of the standard deviation, but generally similar to those of the \FLTR$_{\ICR}$. The \gini values are less high, indicating a more equitable distribution of ranks.
In the same way, we see practically the same values for \FLTR$_{1-\PageRank}$ and \FLTR$_{1-\FLTR}$ (see Tables~\ref{tab:randomThresholds77} and~\ref{tab:randomThresholds88} in the Appendix, respectively) in all metrics. The expansion of influence is predictably restrictive unlike the cases described in the previous paragraph. The values of $\sigma$ are very low and those of the \Gini higher in most cases, compared to the previous ones obtained in the non-complementary version. The exceptions are in \Higgs and \Gnutella, where they previously had high values but are now diminished.

\section{Conclusions}
\label{sec:conclusion}

We have considered the spread of influence in the LT model and the absence of potential values for the threshold defining the individual resistance that is the basis for the model. We have selected to study the behaviour of the \FLTR rank in a popular selection of social networks, and studied different ways to fix the individual threshold of the actors. Our results show that the selection has a clear impact on the ranking.  Surprisingly enough, we have found a abrupt variation of the ranking when setting the thresholds uniformly, being thresholds in the interval $[0.2,0.5]$ the best selection. For the case of random thresholds, a similar phenomenon appears as there are almost no differences on the ranking when the individual thresholds are selected in $(0,0.5]$ than in $(0,1]$. Finally, we have been able to asses that the social networks ranks that provide the best assignments for the individual thresholds are \PageRank and \FLTR.

In this paper, we have focused on the effect of the selection of the influence resistance on the \FLTR ranking. There are many ways to complement this study, for example by including  additional social networks in the analysis (in particular, synthetic social networks). This line of research might provide insight on the best models to study influence spread in the LT model. Another extension will be to study how the threshold selection impacts on other influence spread problems like the influence maximization or the target set selection problems.   

%\printbibliography
\bibliographystyle{alpha}
\bibliography{main}

\newpage
%\newpage
%\pagestyle{plain}
\section*{APPENDIX}
%\label{sec:appendix}

%%%%
%%%% Experimental results when considering uniform thresholds 
%%%% (networks not included in the paper)
%%%%

\begin{table*}[h!]
\centering
\caption{Experimental results when considering uniform thresholds (refinement - continuation).}
\label{tab:uniformThresholds3}
    \centering
    \scalebox{0.9}{
    \begin{tabular}{lrrrlllrlr}
    \cline{2-10}
    &
    \textbf{$\theta$} & 
    \multicolumn{1}{c}{\textbf{$\sigma$}} & 
    \multicolumn{1}{c}{\textbf{\#}} & 
    \multicolumn{1}{c}{\textbf{Gini}} & 
    \multicolumn{1}{c}{\textbf{\topTEN}} & 
    \multicolumn{2}{c}{\textbf{\topTENa}} & 
    \multicolumn{2}{c}{\textbf{\topTENv}} \\
    \cline{2-10}
    \parbox[t]{2mm}{\multirow{11}{*}{\rotatebox[origin=c]{90}{\textbf{\Amazon}}}} 
     & 0.20 &  0.438453 &  1190 &  0.580852 &    0.890206 &  0.904492 & (33486) &  0.904492 & (140309) \\
     & 0.22 &  0.425653 &  1394 &  0.627482 &    0.882755 &  0.894873 & = &  0.894906 & (124701) \\
     & 0.24 &  0.356763 &  1680 &  0.779655 &     0.86286 &  0.871156 & = &   0.871156 & (73847) \\
     & 0.26 &  0.001288 &  1141 &  0.686987 &  0.00642651 &  0.755064 & = &     0.634445 & (745) \\
     & 0.28 &  0.000433 &   944 &  0.631954 &   0.0220418 &   0.74156 & = &     0.572371 & (493) \\
     & 0.30 &  0.000210 &   750 &  0.573468 &  0.00124529 &  0.699662 & = &    0.0274232 & (167) \\
     & 0.32 &  0.000185 &   676 &  0.558115 &  0.00639366 &  0.689819 & = &    0.0190287 & (121) \\
     & 0.34 &  0.000094 &   424 &  0.519286 &   0.0018097 &  0.539665 & = &    0.00560528 & (53) \\
     & 0.36 &  0.000091 &   409 &  0.512088 &  0.00177983 &  0.537644 & = &    0.00549777 & (45) \\
     & 0.38 &  0.000084 &   385 &  0.501672 &  0.00179775 &  0.508196 & = &    0.00421068 & (61) \\
     & 0.40 &  0.000063 &   334 &  0.457212 &  0.00163052 &   0.42713 & = &    0.00321923 & (54) \\
    \cline{2-10}
    \parbox[t]{2mm}{\multirow{16}{*}{\rotatebox[origin=c]{90}{\textbf{\ArXiv}}}} 
     & 0.20 &  0.350389 &  290 &  0.562644 &    0.725486 &  0.740366 & (524) &    0.725677 & (74) \\
     & 0.22 &  0.344483 &  332 &  0.600133 &    0.724723 &  0.739794 & = &    0.725296 & (88) \\
     & 0.24 &  0.282045 &  435 &  0.762620 &   0.0599008 &  0.722434 & = &    0.71881 & (126) \\
     & 0.26 &  0.148570 &  430 &  0.853959 &    0.648417 &  0.670355 & = &   0.656429 & (109) \\
     & 0.28 &  0.090378 &  428 &  0.805350 &   0.0345288 &  0.665395 & = &    0.634681 & (76) \\
     & 0.30 &  0.027497 &  369 &  0.670478 &   0.0328119 &  0.646127 & = &    0.608928 & (38) \\
     & 0.32 &  0.019953 &  342 &  0.640527 &   0.0217474 &   0.64689 & = &   0.0940481 & (40) \\
     & 0.34 &  0.011081 &  260 &  0.595235 &  0.00667684 &  0.526326 & = &   0.0410149 & (27) \\
     & 0.36 &  0.010573 &  255 &  0.589227 &  0.00667684 &  0.528997 & = &   0.0410149 & (33) \\
     & 0.38 &  0.009530 &  240 &  0.578978 &  0.00534147 &  0.498855 & = &   0.0394887 & (27) \\
     & 0.40 &  0.008346 &  215 &  0.568466 &  0.00591377 &  0.450591 & = &   0.0352919 & (23) \\
     & 0.42 &  0.008209 &  215 &  0.565402 &  0.00591377 &  0.446776 & = &   0.0223197 & (22) \\
     & 0.44 &  0.007836 &  205 &  0.560768 &  0.00572301 &  0.430942 & = &   0.0194582 & (21) \\
     & 0.46 &  0.007448 &  197 &  0.555147 &  0.00572301 &   0.41034 & = &   0.0183136 & = \\
     & 0.48 &  0.007343 &  196 &  0.553004 &  0.00476917 &  0.408623 & = &   0.0183136 & (22) \\
     & 0.50 &  0.005950 &  169 &  0.540466 &  0.00267074 &  0.263068 & = &  0.00839374 & (20) \\
    \cline{2-10}
    \parbox[t]{2mm}{\multirow{11}{*}{\rotatebox[origin=c]{90}{\textbf{\ENRON}}}} 
     & 0.20 &  0.399316 &  1603 &  0.651393 &   0.885179 &  0.888423 & (3669) &  0.885179 & (193) \\
     & 0.22 &  0.368746 &  1993 &  0.717240 &   0.876322 &  0.877957 & = &  0.876322 & (267) \\
     & 0.24 &  0.327435 &  2371 &  0.780700 &   0.868336 &  0.873678 & = &  0.868336 & (346) \\
     & 0.26 &  0.260311 &  2801 &  0.841103 &   0.110406 &  0.845416 & = &  0.841655 & (426) \\
     & 0.28 &  0.232640 &  2865 &  0.857943 &   0.102529 &  0.837049 & = &  0.836695 & (456) \\
     & 0.30 &  0.182656 &  3112 &  0.870363 &  0.0612668 &  0.822332 & = &  0.785948 & (522) \\
     & 0.32 &  0.162687 &  3273 &  0.868570 &  0.0528726 &  0.819879 & = &  0.778998 & (478) \\
     & 0.34 &  0.130052 &  3036 &  0.867378 &  0.0407173 &   0.76191 & = &  0.721138 & (446) \\
     & 0.36 &  0.115748 &  3095 &  0.858108 &  0.0410989 &  0.750382 & = &   0.70931 & (404) \\
     & 0.38 &  0.100060 &  3059 &  0.843966 &  0.0391639 &  0.737545 & = &  0.689742 & (364) \\
     & 0.40 &  0.082659 &  2855 &  0.830582 &  0.0394636 &  0.710673 & = &  0.657173 & (340) \\
    \cline{2-10}
    \parbox[t]{2mm}{\multirow{11}{*}{\rotatebox[origin=c]{90}{\textbf{\Epinions}}}} 
     & 0.20 &  0.065796 &  1487 &  0.973600 &  0.0531636 &  0.621714 & (7587) &  0.619025 & (1014) \\
     & 0.22 &  0.057338 &  1541 &  0.972229 &  0.0563133 &  0.621621 & = &   0.618999 & (804) \\
     & 0.24 &  0.046325 &  1590 &  0.967139 &  0.0514108 &  0.621055 & = &   0.618182 & (584) \\
     & 0.26 &  0.032387 &  1412 &  0.957486 &  0.0462051 &  0.612989 & = &   0.609418 & (345) \\
     & 0.28 &  0.024788 &  1428 &  0.941869 &  0.0461129 &  0.613569 & = &   0.608719 & (257) \\
     & 0.30 &  0.018884 &  1387 &  0.924010 &  0.0365977 &   0.61096 & = &   0.604581 & (208) \\
     & 0.32 &  0.016061 &  1397 &  0.913228 &  0.0374017 &  0.610973 & = &    0.60122 & (184) \\
     & 0.34 &  0.009734 &  1146 &  0.886987 &  0.0243941 &   0.58957 & = &   0.574428 & (128) \\
     & 0.36 &  0.008282 &  1134 &  0.878990 &  0.0217715 &  0.589504 & = &    0.11065 & (127) \\
     & 0.38 &  0.005413 &  1147 &  0.863924 &   0.025857 &  0.587462 & = &  0.0967066 & (118) \\
     & 0.40 &  0.003417 &  1137 &  0.855047 &  0.0257647 &   0.57962 & = &  0.0874946 & (119) \\
    \cline{2-10}
    \parbox[t]{2mm}{\multirow{11}{*}{\rotatebox[origin=c]{90}{\textbf{\Wikipedia}}}} 
     & 0.20 &  0.048013 &  228 &  0.917849 &    0.325791 &  0.366831 & (711) &   0.335067 & (180) \\
     & 0.22 &  0.039653 &  282 &  0.907290 &     0.32565 &  0.367112 & = &   0.331553 & (119) \\
     & 0.24 &  0.033243 &  287 &  0.893234 &    0.325088 &  0.366971 & = &    0.327337 & (36) \\
     & 0.26 &  0.025270 &  310 &  0.869475 &     0.32298 &  0.366409 & = &    0.324526 & (35) \\
     & 0.28 &  0.020109 &  291 &  0.846199 &    0.322699 &  0.366409 & = &    0.323963 & (32) \\
     & 0.30 &  0.014252 &  288 &  0.816121 &  0.00309206 &  0.366268 & = &    0.014617 & (30) \\
     & 0.32 &  0.011766 &  277 &  0.803275 &   0.0035137 &  0.366128 & = &   0.0116655 & (28) \\
     & 0.34 &  0.009543 &  261 &  0.789090 &  0.00210822 &  0.365566 & = &  0.00927618 & = \\
     & 0.36 &  0.009013 &  261 &  0.785309 &  0.00210822 &  0.365425 & = &  0.00955727 & = \\
     & 0.38 &  0.008702 &  246 &  0.782338 &  0.00210822 &  0.364863 & = &  0.00801124 & (26) \\
     & 0.40 &  0.007663 &  255 &  0.775823 &  0.00210822 &  0.364582 & = &  0.00674631 & (27) \\
    \cline{2-10}
    \end{tabular}}
\end{table*}

%%%%
%%%% Details on the thresholds assigned to the networks 
%%%% when they are fixed by other centrality measures
%%%%

%\begin{landscape}
\begin{table*}[t!]
\centering
\caption{Details on the thresholds assigned to the networks for \FLTR$_{\mu}$}
%when they are fixed by centrality measure $\mu$.}%, i.e., the so-called FLTR$_{\mu}$.}
\label{tab:randomThresholds4}
    \centering
    \scalebox{1.0}{\vbox{
    \begin{tabular}{l|lccc|clcc}
    \textbf{Network}   &  \textbf{Min} & \textbf{Max} & \textbf{Avg} & \textbf{$\sigma$} &     \textbf{Min} & \textbf{Max} & \textbf{Avg} & \textbf{$\sigma$} \\
    \hline
    \multicolumn{1}{c}{} & \multicolumn{4}{c}{$\mu:$\textbf{\Betweenness}} & \multicolumn{4}{c}{$\mu:$\textbf{\ICR}} \\
    \hline
       \Amazon &  0.0 &      0.0322 &  0.0000 &  0.0002 &  0.0008 &  1.0 &  0.0082 &  0.0286\\
        \ArXiv &  0.0 &      0.0741 &  0.0012 &  0.0039 &  0.0009 &  1.0 &  0.2028 &  0.2450\\
        \Caida &  0.0 &      0.1538 &  0.0001 &  0.0021 &  0.0003 &  1.0 &  0.1378 &  0.1294\\
        \ENRON &  0.0 &      0.1297 &  0.0001 &  0.0018 &  0.0001 &  1.0 &  0.4136 &  0.3124\\
        \Epinions &  0.0 &   0.0655 &  0.0001 &  0.0008 &  0.0001 &  1.0 &  0.1765 &  0.2656\\
        \Gnutella &  0.0 &   0.0028 &  0.0000 &  0.0001 &  0.0763 &  1.0 &  0.1048 &  0.0550\\
        \Higgs &  0.0 &      0.0390 &  0.0000 &  0.0002 &  0.0474 &  1.0 &  0.0561 &  0.0142\\
      \Wikipedia &  0.0\textcolor{white}{000} &    0.0177 &  0.0001 &  0.0005 &  0.0006 &  1.0\textcolor{white}{000} &  0.2290 &  0.3064\\
    \end{tabular}
    \begin{tabular}{l|llll|llll}
    \hline
    \multicolumn{1}{c}{} & \multicolumn{4}{c}{$\mu:$\textbf{\PageRank}} & \multicolumn{4}{c}{$\mu:$\textbf{\FLTR}} \\
%              &  \textbf{Min} & \textbf{Max} & \textbf{Avg} & \textbf{$\sigma$} & \textbf{Min} & \textbf{Max} & \textbf{Avg} & \textbf{$\sigma$} \\
    \hline
       \Amazon &  0.0000 &   0.0002 &  0.0000 &  0.0000 &  0.0000 &  0.0019 &  0.0000 &  0.0000 \\
        \ArXiv &  0.0000 &   0.0014 &  0.0002 &  0.0001 &  0.0006 &  0.0630 &  0.0041 &  0.0060 \\
        \Caida &  0.0000 &   0.0216 &  0.0000 &  0.0003 &  0.0001 &  0.4731 &  0.0072 &  0.0149 \\
        \ENRON &  0.0000 &   0.0114 &  0.0000 &  0.0001 &  0.0001 &  0.6488 &  0.0097 &  0.0343 \\
        \Epinions & 0.0000 &   0.0047 &  0.0000 &  0.0001 &  0.0000 &  0.1014 &  0.0003 &  0.0013 \\
        \Gnutella & 0.0000 &   0.0001 &  0.0000 &  0.0000 &  0.0000 &  0.0043 &  0.0001 &  0.0002 \\
        \Higgs &  0.0000 &   0.0289 &  0.0000 &  0.0001 &  0.0000 &  0.0005 &  0.0000 &  0.0000 \\
      \Wikipedia &  0.0001 &   0.0046 &  0.0001 &  0.0002 &  0.0001 &  0.1524 &  0.0023 &  0.0064 \\
    \hline
    \end{tabular}}}
\end{table*}
%\end{landscape}
%
%%%%
%%%% Thresholds Fixed by Other Centrality Measures 
%%%%
%

\begin{table*}[h!]
\centering
\caption{Experimental results when the thresholds of the nodes are assigned according to 1-\Betweenness.}
\label{tab:randomThresholds55}
    \centering
    \scalebox{1.0}{\vbox{
    \begin{tabular}{lrrrlllrlr}
    \hline
    \textbf{Network} &
    \multicolumn{1}{c}{\textbf{$\sigma$}} & 
    \multicolumn{1}{c}{\textbf{\#}} & 
    \multicolumn{1}{c}{\textbf{Gini}} & 
    \multicolumn{1}{c}{\textbf{\topTEN}} & 
    \multicolumn{2}{c}{\textbf{\topTENa}} & 
    \multicolumn{2}{c}{\textbf{\topTENv}} \\
    \hline
       \Amazon &  0.000168 &   280 &  0.731851 &  0.000680875 &  0.121023 & (33486) &  0.000752547 & (32) \\
        \ArXiv &  0.003687 &   120 &  0.514478 &   0.00190767 &    0.102823 & (524) &    0.0028615 & (15) \\
        \Caida &  0.002197 &   219 &  0.679585 &  0.000377715 &   0.275279 & (2647) &    0.0123513 & (22) \\
        \ENRON &  0.010085 &  1533 &  0.721958 &  0.000272539 &   0.118554 & (3669) &   0.0063229 & (161) \\
     \Epinions &  0.000521 &   450 &  0.761468 &   0.00322882 &   0.169177 & (7587) &   0.00769646 & (48) \\
     \Gnutella &  0.000070 &    52 &  0.545161 &   0.00015978 &  0.0999904 & (6258) &   0.000111846 & (7) \\
        \Higgs &  0.000006 &    50 &  0.203628 &  0.000128659 &  0.106479 & (25649) &    8.5773e-05 & (6) \\
    \Wikipedia &  0.005972 &   239 &  0.761551 &   0.00154603 &    0.103162 & (711) &    0.0037948 & (25) \\
    \hline
    \end{tabular}}}
%\end{table*}
%

\bigskip
%
%\begin{table*}[hp!]
\centering
\caption{Experimental results when the thresholds of the nodes are assigned according to 1-\ICR.}
\label{tab:randomThresholds66}
    \centering
    \begin{tabular}{lrrrlllrlr}
    \hline
    \textbf{Network} &
    \multicolumn{1}{c}{\textbf{$\sigma$}} & 
    \multicolumn{1}{c}{\textbf{\#}} & 
    \multicolumn{1}{c}{\textbf{Gini}} & 
    \multicolumn{1}{c}{\textbf{\topTEN}} & 
    \multicolumn{2}{c}{\textbf{\topTENa}} & 
    \multicolumn{2}{c}{\textbf{\topTENv}} \\
    \hline
       \Amazon &  0.000031 &  204 &  0.368671 &  0.000788382 &  0.188737 & (33486) &  0.00104819 & (26) \\
        \ArXiv &  0.042332 &  315 &  0.641623 &    0.0925219 &    0.232354 & (524) &    0.121709 & (45) \\
        \Caida &  0.157404 &  385 &  0.230893 &     0.373749 &   0.636525 & (2647) &    0.383645 & (40) \\
        \ENRON &  0.209299 &  838 &  0.097727 &     0.707729 &   0.728061 & (3669) &    0.707729 & (88) \\
     \Epinions &  0.188269 &  137 &  0.331004 &     0.400071 &   0.531504 & (7587) &    0.400071 & (15) \\
     \Gnutella &  0.000177 &  120 &  0.692785 &   0.00375483 &    0.91455 & (6258) &  0.00452178 & (14) \\
        \Higgs &  0.000006 &   52 &  0.214404 &  0.000362586 &  0.124737 & (25649) &  0.000304104 & (6) \\
    \Wikipedia &  0.056027 &  173 &  0.451150 &     0.116796 &    0.214758 & (711) &    0.119606 & (18) \\
    \hline
    \end{tabular}
%\end{table*}
%

\bigskip
%
%\begin{table*}[hp!]
\centering
\caption{Experimental results when the thresholds of the nodes are assigned according to 1-\PageRank.}
\label{tab:randomThresholds77}
    \centering
    \begin{tabular}{lrrrlllrlr}
    \hline
    \textbf{Network} &
    \multicolumn{1}{c}{\textbf{$\sigma$}} & 
    \multicolumn{1}{c}{\textbf{\#}} & 
    \multicolumn{1}{c}{\textbf{Gini}} & 
    \multicolumn{1}{c}{\textbf{\topTEN}} & 
    \multicolumn{2}{c}{\textbf{\topTENa}} & 
    \multicolumn{2}{c}{\textbf{\topTENv}} \\
    \hline
       \Amazon &  0.000031 &   201 &  0.366149 &  0.000791368 &  0.189304 & (33486) &  0.00109896 & (27) \\
        \ArXiv &  0.004110 &   132 &  0.513894 &   0.00247997 &    0.163869 & (524) &   0.0040061 & (16) \\
        \Caida &  0.012823 &  1169 &  0.693383 &    0.0964306 &   0.462285 & (2647) &   0.260283 & (132) \\
        \ENRON &  0.012020 &  1776 &  0.710027 &    0.0231658 &   0.334514 & (3669) &   0.145945 & (183) \\
     \Epinions &  0.001183 &   807 &  0.827930 &    0.0144045 &   0.334256 & (7587) &   0.0379288 & (86) \\
     \Gnutella &  0.000172 &   111 &  0.689222 &   0.00594382 &   0.672307 & (6258) &  0.00640718 & (12) \\
        \Higgs &  0.000006 &    52 &  0.214376 &  0.000354788 &  0.124737 & (25649) &  0.000304104 & (6) \\
    \Wikipedia &  0.006101 &   235 &  0.762919 &   0.00196767 &    0.116655 & (711) &  0.00477864 & (27) \\
    \hline
    \end{tabular}
%\end{table*}
%

\bigskip
%
%\begin{table*}[hp!]
\centering
\caption{Experimental results when the thresholds of the nodes are assigned according to 1-\FLTR.}
\label{tab:randomThresholds88}
    \centering
    \begin{tabular}{lrrrlllrlr}
    \hline
    \textbf{Network} &
    \multicolumn{1}{c}{\textbf{$\sigma$}} & 
    \multicolumn{1}{c}{\textbf{\#}} & 
    \multicolumn{1}{c}{\textbf{Gini}} & 
    \multicolumn{1}{c}{\textbf{\topTEN}} & 
    \multicolumn{2}{c}{\textbf{\topTENa}} & 
    \multicolumn{2}{c}{\textbf{\topTENv}} \\
    \hline
       \Amazon &  0.000031 &   201 &  0.366149 &  0.000791368 &  0.189304 & (33486) &  0.00109896 & (27) \\
        \ArXiv &  0.004110 &   132 &  0.513894 &   0.00247997 &    0.163869 & (524) &   0.0040061 & (16) \\
        \Caida &  0.012823 &  1169 &  0.693383 &    0.0964306 &   0.462285 & (2647) &   0.260283 & (132) \\
        \ENRON &  0.012102 &  1780 &  0.710349 &    0.0187507 &   0.334514 & (3669) &   0.145999 & (185) \\
     \Epinions &  0.001183 &   807 &  0.827930 &    0.0144045 &   0.334256 & (7587) &   0.0379288 & (86) \\
     \Gnutella &  0.000172 &   111 &  0.689222 &   0.00594382 &   0.672307 & (6258) &  0.00640718 & (12) \\
        \Higgs &  0.000006 &    52 &  0.214376 &  0.000354788 &  0.124737 & (25649) &  0.000304104 & (6) \\
    \Wikipedia &  0.006101 &   235 &  0.762919 &   0.00196767 &    0.116655 & (711) &  0.00477864 & (27) \\
    \hline
    \end{tabular}
\end{table*}

\end{document}